\documentclass[a4paper]{PoS}

\usepackage[caption=false]{subfig}
\usepackage{amsmath}

\title{Radiative corrections in Dalitz decays of $\pi^0$, $\eta$ and $\eta^\prime$ mesons}

\ShortTitle{Radiative corrections in Dalitz decays of $\pi^0$, $\eta$ and $\eta^\prime$ mesons}

\author{\speaker{Tom\'a\v{s} Husek}\\
        IFIC, Universitat de Val\`encia -- CSIC, Apt.\ Correus 22085, E-46071 Val\`encia, Spain\\
        E-mail: \email{thusek@ific.uv.es}}

\abstract{
We briefly summarize current experimental and theoretical results on the two important processes of the low-energy hadron physics involving neutral pions, the rare decay $\pi^0\to e^+e^-$ and the Dalitz decay of $\pi^0$, and provide a new value for the ratio $R={\Gamma(\pi^0\to e^+e^-\gamma(\gamma))}/{\Gamma(\pi^0\to\gamma\gamma)}=11.978(6)\times10^{-3}$, which is by two orders of magnitude more precise than the current PDG average.
This value is obtained using the complete set of the next-to-leading-order radiative corrections in the QED sector, and incorporates up-to-date values of the $\pi^0$-transition-form-factor slope.
The ratio $R$ translates into the branching ratios of the two main $\pi^0$ decay modes: $\mathcal{B}(\pi^0\to\gamma\gamma)=98.8131(6)\,\%$ and $\mathcal{B}(\pi^0\to e^+e^-\gamma(\gamma))=1.1836(6)\,\%$.
Furthermore, we present the complete set of radiative corrections to the Dalitz decays $\eta^{(\prime)}\to\ell^+\ell^-\gamma$ beyond the soft-photon approximation, i.e.\ over the whole range of the Dalitz plot and with no restrictions on the energy of a radiative photon.
The corrections inevitably depend on the $\eta^{(\prime)}\to\gamma^*\gamma^{(*)}$ transition form factors.
For the singly-virtual transition form factor appearing e.g.\ in the bremsstrahlung correction, recent dispersive calculations are used.
For the one-photon-irreducible contribution at the one-loop level (for the doubly-virtual form factor), we use a vector-meson-dominance-inspired model while taking into account the $\eta$-$\eta^{\prime}$ mixing.
}

\FullConference{}

\def\cuteps{\epsilon}

\makeatletter
\providecommand*{\diff}%
  {\@ifnextchar^{\DIfF}{\DIfF^{}}}
\def\DIfF^#1{%
  \mathop{\mathrm{\mathstrut d}}%
    \nolimits^{#1}\gobblespace}
\def\gobblespace{%
    \futurelet\diffarg\opspace}
\def\opspace{%
    \let\DiffSpace\!%
    \ifx\diffarg(%
      \let\DiffSpace\relax
     \else
      \ifx\diffarg[%
	\let\DiffSpace\relax
      \else
	\ifx\diffarg\{%
	  \let\DiffSpace\relax
	\fi\fi\fi\DiffSpace}

\begin{document}

\section{Introduction}
\label{intro}

The rare decay of the neutral pion, i.e.\ the process $\pi^0\to e^+e^-$, is loop- and helicity-suppressed compared to the two-photon decay.
This significant suppression makes it potentially sensitive to effects of new physics.
That is why it drew attention of theorists during last years due to the precise measurement of its branching ratio done by KTeV experiment at Fermilab~\cite{Abouzaid:2006kk}:
\begin{equation}
\mathcal{B}(\pi^0\to e^+e^-(\gamma),\,x>0.95)\big|_\text{KTeV}=(6.44\pm0.25\pm0.22)\times10^{-8}\,.
\end{equation}
Subsequent comparison with the Standard Model prediction was performed~\cite{Dorokhov:2007bd} and interpreted as a 3.3\,$\sigma$ discrepancy between theory and experiment.
One could immediately think that this might be a sign for new physics.
However, a more conventional solution should be sought first.

In what follows we will investigate in detail the radiative corrections for neutral-pion decays in general.
Another possibility could be introducing an advanced model for the $\pi^0$ electromagnetic transition form factor.
In Ref.~\cite{Husek:2015wta} we discussed this option on the case of a novel two-hadron saturation (THS) model, which belongs to the family of large-$N_c$-motivated resonance-saturation models.
The THS model is phenomenologically successful and by construction surpasses other related models (VMD, LMD), since it simultaneously satisfies many LO high-energy constraints (e.g.\ the operator product expansion (OPE)~\cite{Knecht:2001xc,Husek:2015wta} and the Brodsky--Lepage scaling limit~\cite{Lepage:1979zb,Lepage:1980fj}).
Any further discussion is though beyond the scope of this contribution.
Let us just briefly mention that electromagnetic transition form factors represent also valuable inputs to the anomalous magnetic moment ($g-2$) of the muon.
For instance, within the THS model the pion-pole contribution to the hadronic light-by-light piece of the muon $g-2$ reads $a_\mu^{\text{LbL};\pi^0}=75(5)\times10^{-11}$.

Two-loop virtual radiative corrections for the $\pi^0$ rare decay were calculated in Ref.~\cite{Vasko:2011pi} and the bremsstrahlung beyond the soft-photon approximation was discussed in Ref.~\cite{Husek:2014tna}.
The final next-to-leading-order (NLO) correction was found to be $\delta^\text{NLO}(0.95)=-5.5(2)\,\%$, which differs significantly from the previous approximate results also used in Ref.~\cite{Abouzaid:2006kk}.
When the exactly calculated radiative corrections are taken into account, the original discrepancy reduces down to the inconclusive 2\,$\sigma$ level or below~\cite{Husek:2014tna,Husek:2015wta}.

\section{Neutral-pion Dalitz decay}

We see that correct incorporation of radiative corrections is crucial in order to provide relevant experimental results.
In the rare-pion-decay search performed by the KTeV experiment, the Dalitz decay was used as the normalization channel.
It is thus important to have the radiative corrections for also this decay under control, among others in order to correctly extract valuable information about the singly-virtual transition form factor from experiment.

Radiative corrections to the total decay rate were first (numerically) addressed by Joseph~\cite{Joseph:1960zz}.
A pioneering study of the corrections to the differential decay rate was done by Lautrup and Smith~\cite{Lautrup:1971ew}, although only in the soft-photon approximation.
This was extended later in the classical work of Mikaelian and Smith~\cite{Mikaelian:1972yg}, where the corrections to the Dalitz plot in the form of a table of values were presented.

The new investigation~\cite{Husek:2015sma} of this topic was motivated by needs of NA48/NA62 experiments at CERN, the aim of which, among other goals, was to measure the slope $a_\pi$ of the singly-virtual transition form factor $\mathcal{F}_{\pi^0\gamma^*\gamma^*}(0,q^2)$\,.
Unlike before, we took into account the one-photon-irreducible ($1\gamma$IR) contribution, which was, due to inappropriate assumptions and arguments based on the Low's theorem~\cite{Low:1958sn,Adler:1966gc,Pestleau:1967snm}, considered to be negligible and left out in the previous works; see also Refs.~\cite{Mikaelian:1972yg,Lambin:1985sb}.
The exact calculation though shows it to be significant~\cite{Tupper:1983uw,Tupper:1986yk,Kampf:2005tz,Husek:2015sma}.
Additionally, no approximation regarding masses of the particles involved was used during the calculation, so the results are applicable for related decays like $\eta\to \ell^+\ell^-\gamma$, although additional prudence and treatment is required.
Finally, the \verb!C++! code was developed, which returns the radiative correction for any given kinematically allowed point.
This code became a part of the Monte Carlo event generator in the NA62 experiment.
Subsequently, 1.1$\times10^6$ fully reconstructed $\pi^0$ Dalitz decays were analyzed with the result $a_\pi^\text{NA62}=3.68(57)\,\%$~\cite{TheNA62:2016fhr}.
The current PDG value $a_\pi^\text{PDG}=3.35(31)$~\cite{Tanabashi:2018oca} is dominated by two inputs: the above NA62 precise time-like-region result and the value provided by the CELLO collaboration ($a_\pi^\text{CELLO}=3.26(37)\,\%$)~\cite{Behrend:1990sr} using the (model-dependent) extrapolation from the space-like region.

Having at hand a relatively good knowledge of the form-factor shape (given sufficiently in the case of the neutral pion by the form-factor slope $a_\pi$) and the complete set of NLO QED radiative corrections $\delta(x,y)$, we can determine very precisely and reliably the following ratio~\cite{Husek:2018qdx}:
\begin{equation}
R
=\frac{\Gamma(\pi^0\to e^+e^-\gamma(\gamma))}{\Gamma(\pi^0\to\gamma\gamma)}
\simeq\frac{\alpha}{\pi}
\iint
\,(1+a_\pi x)^2
(1+\delta(x,y))\\
\frac{(1-x)^3}{4x}\left[1+y^2+\frac{4m_e^2}{M_\pi^2x}\right]\diff x\diff y\,.
\label{eq:R}
\end{equation}
Choosing $a_\pi^\text{univ}\equiv3.55(70)\,\%$ which covers a whole interval of numerical values suggested by various experiments and theoretical predictions, we find $R=11.978(5)(3)\times10^{-3}$~\cite{Husek:2018qdx}.
The former uncertainty stands for the form-factor effects and the latter for neglecting the higher-order corrections.
This value represents a significant improvement (by two orders of magnitude) to the current PDG-based value of $R$, which might be further used in theoretical predictions and experimental analyses.
Using the fact that branching ratios should sum up to 1, the ratio $R$ translates into the branching ratios of the two main $\pi^0$ decay modes: $\mathcal{B}(\pi^0\to\gamma\gamma)=98.8131(6)\,\%$ and $\mathcal{B}(\pi^0\to e^+e^-\gamma(\gamma))=1.1836(6)\,\%$~\cite{Husek:2018qdx}.

Note that the above precise determination of $R$ and related quantities is possible due to the fact that the peculiar form-factor normalization $\mathcal{F}_{\pi^0\gamma^*\gamma^*}(0,0)$ drops out in $R$ and the form-factor dependence is then solely represented by its shape.
In the neutral-pion-Dalitz-decay case, the transfered momentum squared is significantly (kinematically) limited with the linear expansion being a very good approximation.
The slope parameter $a_\pi$ then constitutes the only relevant parameter from the low-energy QCD sector.
Allowing for such a high (20\,\%) uncertainty on $a_\pi^\text{univ}$ is then the consequence of the smallness of the slope and of the strong suppression of the region $x\simeq1$ where the $a_\pi x$ term matters; cf.\ Eq.~(\ref{eq:R}).

\section{Dalitz decays of $\eta^{(\prime)}$ mesons}

\begin{figure}[!t]
\vspace{-5mm}
%
%
\centering
\subfloat[][]{
\includegraphics[width=0.22\columnwidth]{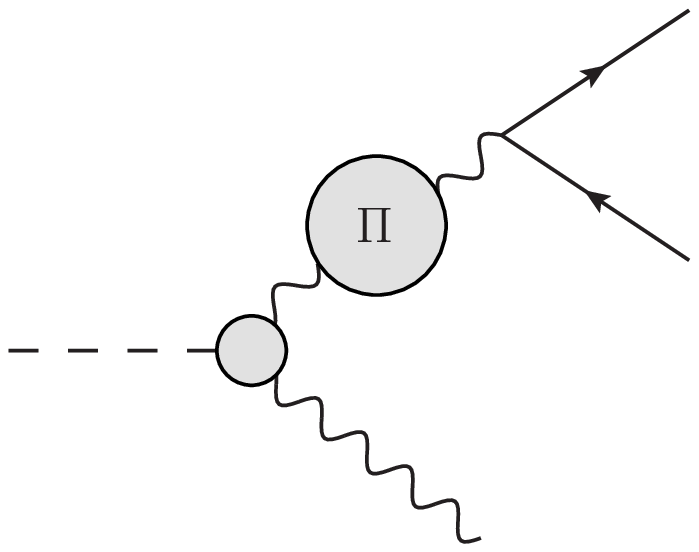}
\label{fig:virta}
}
\hspace{1cm}
\subfloat[][]{
\includegraphics[width=0.22\columnwidth]{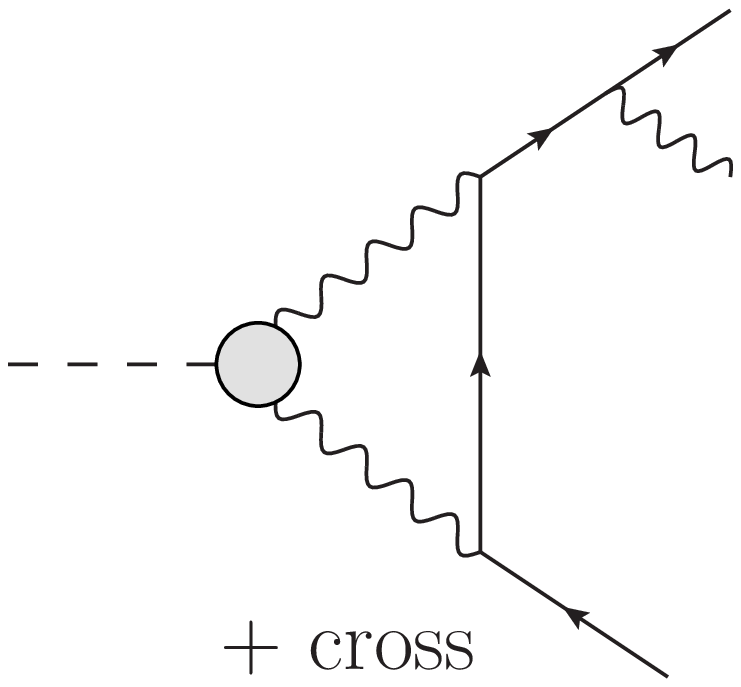}
\label{fig:1gIRa}
}
\hspace{1cm}
\subfloat[][]{
\includegraphics[width=0.22\columnwidth]{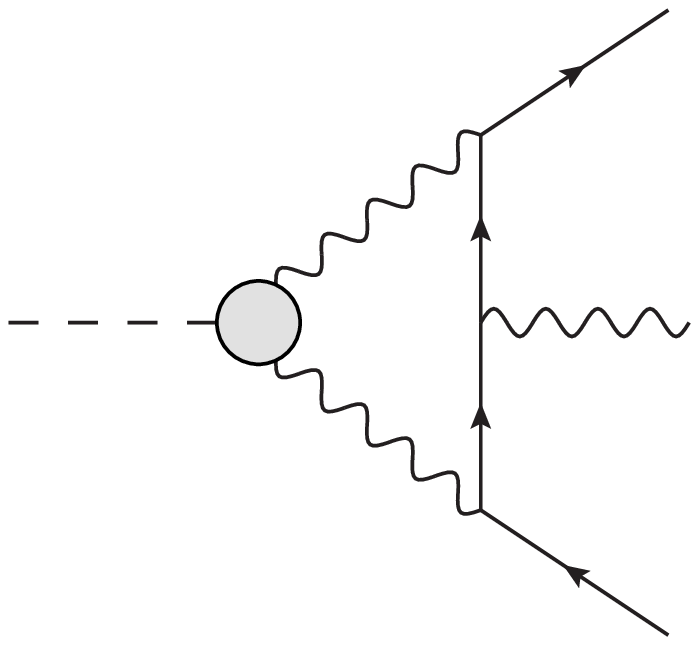}
\label{fig:1gIRb}
}

\subfloat[][]{
\includegraphics[width=0.22\columnwidth]{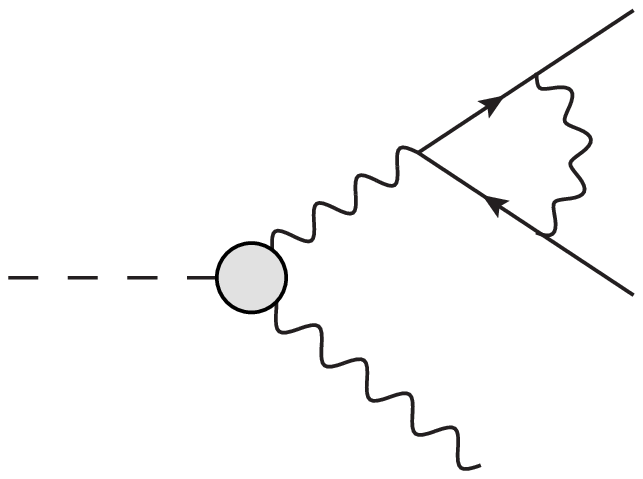}
\label{fig:virtb}
}
\hspace{1cm}
\subfloat[][]{
\includegraphics[width=0.52\columnwidth]{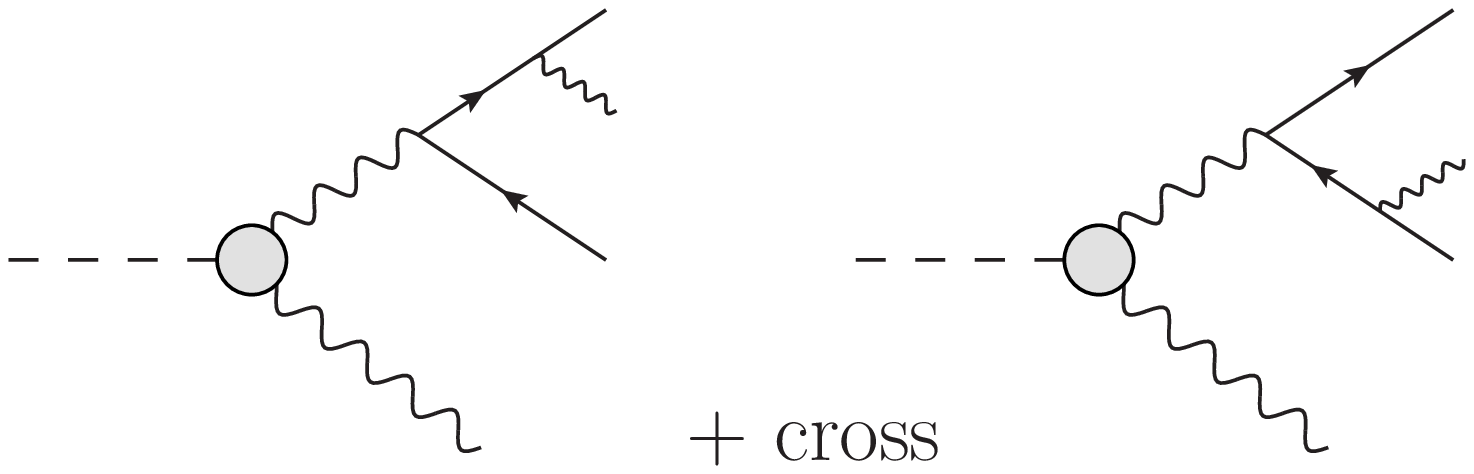}
\label{fig:BS}
}
\caption{
\label{fig:diagrams}
NLO QED radiative corrections to the Dalitz decay $P\to \ell^+\ell^-\gamma$: a) vacuum polarization insertion, b) correction to the QED vertex, c) \& d) one-loop one-photon-irreducible contributions, e) bremsstrahlung.
Note that `cross' in figure (c) corresponds to a diagram where the photon is emitted from the outgoing positron line and `cross' in figure (e) stands for the diagrams with outgoing photons interchanged.
}
\end{figure}

Unlike in the neutral-pion case, due to the higher $\eta^{(\prime)}$ rest masses the $\eta^{(\prime)}$ Dalitz decays do not belong to those with the highest branching ratio: the hadronic decay channels are open.
Nevertheless, studying the Dalitz decays provides a way to access the electromagnetic transition form factors and consequently information about the structure of the related mesons.

Na\"ive radiative corrections for the $\eta\to e^+e^-\gamma$ process were published in Ref.~\cite{Mikaelian:1972jn}: compared to the earlier work~\cite{Mikaelian:1972yg}, only the numerical value of the physical mass of the decaying pseudoscalar was changed.
Ref.~\cite{Husek:2017vmo} completes the list of the NLO corrections in the QED sector and improves the previous approach~\cite{Mikaelian:1972jn}.
Compared to~\cite{Mikaelian:1972jn}, which relates only to the case of the $\eta\to e^+e^-\gamma$ decay, we took into account muon loops and hadronic corrections as a part of the vacuum-polarization contribution, 1$\gamma$IR contribution at one-loop level, higher-order final-state-lepton-mass corrections and form-factor effects.
Moreover, we provide a complete systematic study of the NLO radiative corrections to the differential decay widths to three additional processes including $\eta^\prime$ decays: $\eta\to \mu^+\mu^-\gamma$, $\eta^\prime\to e^+e^-\gamma$ and $\eta^\prime\to \mu^+\mu^-\gamma$.
Sufficiently dense tables of values suitable for interpolation are submitted together with Ref.~\cite{Husek:2017vmo} in a form of ancillary files.

In the case of $\eta$ decays, we could conveniently and extensively draw from the previous work~\cite{Husek:2015sma}, which governed the neutral-pion Dalitz decay, since it was already written in a sufficiently general way.
What really brings the current topic to a different level of difficulty is a desire to tackle the radiative corrections for the $\eta^\prime$ decays.
The resulting framework of Ref.~\cite{Husek:2017vmo} is, of course, directly applicable for the $\eta$ and $\pi^0$ cases.
Let us only mention that the numerical results obtained for the $\pi^0$ Dalitz decay using the new framework are indeed compatible with the form-factor-slope correction suggested at the end of Section V of Ref.~\cite{Husek:2015sma}.
There is thus no particular need to use this generalized framework for the pion case: one gains a correction to the correction at the level of 1\,\%.

Let us shortly discuss the subtleties and difficulties which one encounters and needs to deal with when facing the Dalitz decays of $\eta^{(\prime)}$ mesons and associated NLO radiative corrections and which are mainly driven by the properties of the $\eta^\prime$ meson.
First, it is the higher rest mass, which in the case of $\eta$ is above the muon-pair production threshold and in the case of $\eta^\prime$ even above the lowest-lying resonances $\rho$ and $\omega$, the former of which is a broad resonance in $\pi\pi$ scattering.
This is connected to the fact that the form-factor effects are not negligible as they were in the pion case: the form factor cannot be scaled out anymore and its particular model needs to be taken into account.
We then need to distinguish between two separate cases briefly discussed in next subsections.

\subsection{Bremsstrahlung}

Similarly to the leading-order decay width, in the case of the bremsstrahlung correction the singly-virtual transition form factor appears; the associated diagrams are shown in Fig.~\ref{fig:BS}.
The calculation of this contribution includes integration over angles and energies of the bremsstrahlung photon.
For these integrals to be well-defined in order to obtain reasonable results, including the width of the lowest-lying vector-meson resonances becomes necessary.
Due to the fact that such a calculation will be unavoidably sensitive to the width of the broad $\rho$ resonance, we have decided to incorporate the recent dispersive calculations~\cite{Hanhart:2013vba,Hanhart:2016pcd}.
In the K\"all\'en--Lehmann spectral representation, the form factor has the following form:
\begin{equation}
\frac{\mathcal{F}(q^2)}{\mathcal{F}(0)}
\simeq1-q^2\int_{4m_{\pi}^2}^{\Lambda^2}
\frac{\mathcal{A}(s)\diff s}{q^2-s+i\cuteps}\,,
\end{equation}
where we have used a common spectral density function
\begin{equation}
\mathcal{A}(s)
=\frac1\pi\frac{w_\omega M_\omega\Gamma_\omega}{(s-M_\omega^2)^2+(M_\omega\Gamma_\omega)^2}
+\frac1\pi\frac{w_\phi M_\phi\Gamma_\phi}{(s-M_\phi^2)^2+(M_\phi\Gamma_\phi)^2}
+\frac\kappa{96\pi^2F_\pi^2}
\left[1-\frac{4m_\pi^2}s\right]^{3/2}P(s)|\Omega(s)|^2\,.
\end{equation}
Above, $P(s)$ is a polynomial and $\Omega(s)$ the Omn\`es function, a dispersive tool incorporating pion rescattering.

\subsection{One-photon-irreducible correction}

In the case of the 1$\gamma$IR correction (for diagrams see Figs.~\ref{fig:1gIRa} and \ref{fig:1gIRb}), one needs to take into account the doubly-virtual transition form factor beyond effective approach.
Since the quark content of the $\eta^{(\prime)}$ physical states is not equal to the $\text{U}(3)$ isoscalar states, there is a mixing between $\eta$ and $\eta^\prime$ mesons.
In the quark-flavor basis~\cite{Feldmann:1998sh,Escribano:2005qq},
%
$j^{\ell}\equiv\frac i2\big[\bar u\gamma_5 u+\bar d\gamma_5 d\big], j^\text{s}\equiv\frac i{\sqrt2}\big[{\bar s\gamma_5 s}\big],$
%
this mixing occurs (for $A\in\{\ell,\text{s}\}$) among the states $|\eta^A\rangle$ defined as $\langle 0|j^A|\eta^B\rangle=B_0F_\pi f_A\delta^{AB}$ together with the orthonormality relation $\langle\eta^A|\eta^B\rangle=\delta^{AB}$\,.
In the quark-flavor basis, the mixing can be written as
\begin{align}
|\eta\rangle&=\cos\phi\,|\eta^\ell\rangle-\sin\phi\,|\eta^\text{s}\rangle\,,\label{eq:etamixing}\\
|\eta^{\prime}\rangle&=\sin\phi\,|\eta^\ell\rangle+\cos\phi\,|\eta^\text{s}\rangle\,.
\label{eq:etaprimemixing}
\end{align}
We do not expect any substantial dependence of the resulting corrections on the vector-meson decay widths and we use a simple VMD-inspired model, which incorporates the strange-flavor content of $\eta^{(\prime)}$ mesons and the $\eta$-$\eta^\prime$ mixing.
In the case of the $\eta$ meson, the form factor reads
\begin{equation}
\begin{split}
&e^2\mathcal{F}_{\eta\gamma^*\gamma^*}^\text{VMD}(p^2,q^2)\\
&=-\frac{N_\text{c}}{8\pi^2 F_\pi}
\frac{2e^2}{3}
\left[
\frac53\frac{\cos\phi}{f_\ell}\,\frac{M_{\omega/\rho}^4}{(p^2-M_{\omega/\rho}^2)(q^2-M_{\omega/\rho}^2)}
-\frac{\sqrt2}3\frac{\sin\phi}{f_\text{s}}\,\frac{M_\phi^4}{(p^2-M_\phi^2)(q^2-M_\phi^2)}
\right].
\end{split}
\label{eq:FFVMD}
\end{equation}
Finally, let us mention that the approach used in Ref.~\cite{Husek:2017vmo} is rather general and allows to simply evaluate the contribution to the $1\gamma$IR radiative corrections for the whole family of analytic resonance-saturation models, in which $\mathcal{F}_{P\gamma^*\gamma^*}$ is a rational function with vector-meson poles.

\subsection{Virtual corrections and the photon self-energy}

In the pion case, the vacuum polarization (see Fig.~\ref{fig:virta} for the associated contribution) was dominated by the electron loop.
It turns out though that in the high-invariant-mass region of the photon propagator the hadronic effects become significant, which should be taken into account for the $\eta^{(\prime)}$ decays.
Thus, in general, we shall deal with the photon self-energy in the form $\Pi(s)=\Pi_\text{L}(s)+\Pi_\text{H}(s)$.
For the lepton loops (electrons and muons) we take
\begin{equation}
\Pi_\text{L}(M_P^2x)
=\frac{\alpha}{\pi}\sum_{\ell^\prime=e,\mu}
\left\{\frac89-\frac{\beta_{\ell^\prime}^2(x)}3+\left(1-\frac{\beta_{\ell^\prime}^2(x)}3\right)\frac{\beta_{\ell^\prime}(x)}2\log\left[-\gamma_{\ell^\prime}(x)+i\cuteps\right]\right\},
\end{equation}
where $\beta_{\ell^\prime}(x)\equiv\sqrt{1-{\nu_{\ell^\prime}^2}/{x}}$ and $\gamma_{\ell^\prime}(x)\equiv[1-\beta_{\ell^\prime}(x)]/[1+\beta_{\ell^\prime}(x)]$, with $\nu_{\ell^\prime}\equiv2m_{\ell^\prime}/M_P$ and $x\in[\nu_\ell^2,1]$ ($\ell$ being the final-state lepton in the $\eta^{(\prime)}\to\ell^+\ell^-\gamma$ decay).
The hadronic contribution to the photon self-energy can be expressed via a dispersive integral~\cite{Jegerlehner:1985gq}
\begin{equation}
\Pi_\text{H}(s)
=-\frac{s}{4\pi^2\alpha}\int_{4m_\pi^2}^{\infty}\frac{\sigma_\text{H}(s^\prime)\diff s^\prime}{s-s^\prime+i\cuteps}\,.
\label{eq:PiH}
\end{equation}
Here, $\sigma_\text{H}$ is the total cross section of the $e^+e^-$ annihilation into hadrons and is related to the ratio $R(s)$ as $\sigma_\text{H}(s)=(4\pi\alpha^2/3s)R(s)$.

For completeness, let us revise at this point what we mean by the virtual correction.
In agreement with Eq.~(16) in Ref.~\cite{Husek:2015sma}, we use
\begin{equation}
\delta^\text{virt}(x,y)
=\frac1{|1+\Pi(M_P^2x)|^2}-1+2\operatorname{Re}\left\{F_1(x)+\frac{2F_2(x)}{1+y^2+\frac{\nu_\ell^2}{x}}\right\},
\label{eq:dvirt}
\end{equation}
where $\Pi(s)$ contains not only electron and muon loops, but also the whole hadronic contribution.
For the form factors $F_1$ and $F_2$ arising from the diagram depicted in Fig.~\ref{fig:virtb} the reader is referred to seek the expressions in Ref.~\cite{Husek:2015sma}.

\section{Results and conclusions}

\begin{figure}[t]
\centering
\begin{minipage}{.48\textwidth}
\includegraphics[width=\columnwidth]{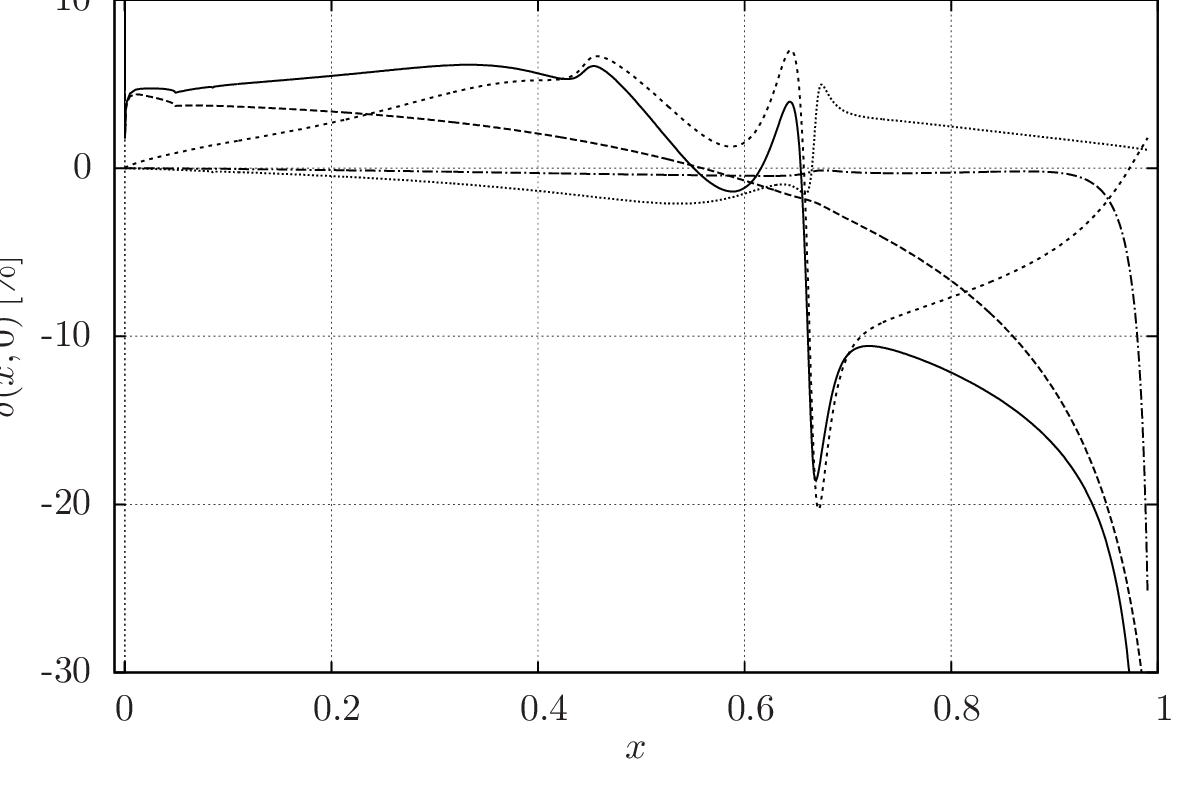}
\caption{
The overall NLO correction $\delta(x, 0)$ (solid line) in comparison to its constituents for the decays $\eta^{(\prime)}\to \ell^+\ell^-\gamma$.
The correction \`a la Mikaelian and Smith~\cite{Mikaelian:1972jn} is depicted as a dashed line.
The hadronic contribution to the virtual radiative corrections is shown as a small-spaced dotted line.
The form-factor-model-dependent part of the bremsstrahlung correction is shown as a large-spaced dotted line.
The one-photon-irreducible contribution is then shown as a dash-dot line.
}
\label{fig:deltax0}
\end{minipage}
\hspace{2mm}
\begin{minipage}{.48\textwidth}
\includegraphics[width=\columnwidth]{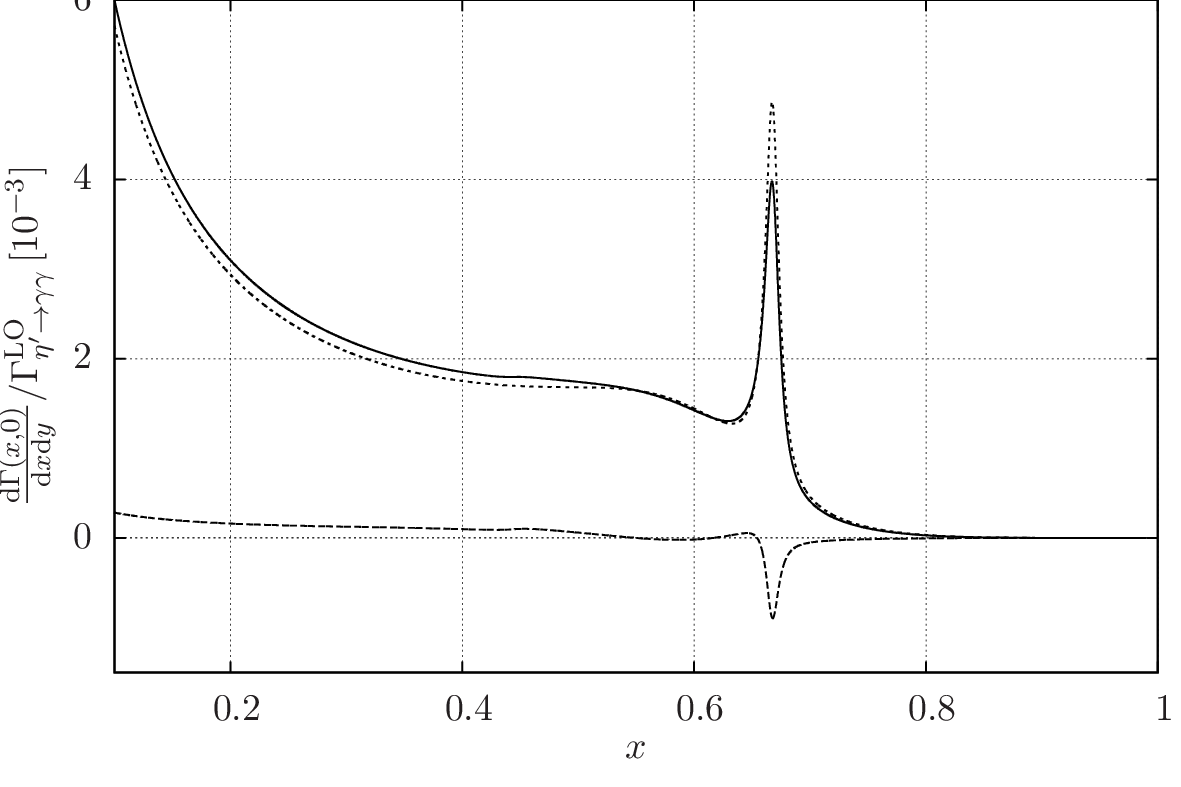}
\caption{
The two-fold differential decay widths $\mathrm{d}\Gamma(x,0)$ at NLO (solid line) and its constituents for the $\eta^{(\prime)}\to\ell^+\ell^-\gamma$ decays.
The LO differential decay width for $y=0$ is shown as a dotted line.
The corresponding NLO contribution to the differential decay width is represented by a dashed line.
\vspace{1.79cm}
}
\label{fig:dGammaNLO}
\end{minipage}
\end{figure}

Examples of the results on the $\eta^\prime\to e^+e^-\gamma$ decay, where the radiative corrections are most significant, are shown in Figs.~\ref{fig:deltax0} and \ref{fig:dGammaNLO}.
Obviously, the form-factor effects are considerable.
The wiggles, which appear only in the $\eta^\prime$ case, then contribute in such a way that they change the shape of the resonance peaks and make them smaller.
In particular, Fig.~\ref{fig:dGammaNLO}c shows that the height of the $\omega$ peak is considerably influenced by the radiative corrections.
This might be interesting for the extraction of $\omega$ properties or of the $\omega$-$\eta^\prime$ interplay.
One might deduce such information from $\eta^\prime\to\omega\gamma\to e^+e^-\gamma$ or from $\eta^\prime\to\omega\gamma\to\pi^+\pi^-\pi^0\gamma$.
It can be expected that the radiative corrections are different for these two decay branches.
Ignoring such radiative corrections in the analyses of these decays might lead to contradictory conclusions.

The complete sets of NLO radiative corrections in the QED sector for the discussed decays --- the Dalitz decays of $\pi^0$~\cite{Husek:2015sma}, $\eta$ and $\eta^\prime$~\cite{Husek:2017vmo} and the rare decay $\pi^0\to e^+e^-$~\cite{Vasko:2011pi,Husek:2014tna} --- are now available and their use in future experimental analyses should be essential.
Note also that in Refs.~\cite{Husek:2015sma,Husek:2017vmo} we study fully inclusive radiative corrections, i.e.\ no momentum or angular cuts on the additional bremsstrahlung photon(s) are applied.

\section*{Acknowledgments}
This contribution is based on collaboration with K.\ Kampf, J. Novotn\'y, S.\ Leupold and E.\ Goudzovski.
Additionally, we would like to thank P. Sanchez-Puertas for providing us with the code to calculate the muon $g-2$ related quantity $a_\mu^{\text{LbL};\pi^0}$.

The work was supported in part by
the Agencia Estatal de Investigaci\'on (AEI, ES) and the European Regional Development Fund (ERDF, EU) [Grants No.\ FPA2014-53631-C2-1-P, FPA2017-84445-P and SEV-2014-0398],
by Generalitat Valenciana [Grant No.\ PROMETEO/2017/053],
by the Czech Science Foundation grant GA\v{C}R 18-17224S and by the ERC starting grant 336581 ``KaonLepton''.


\providecommand{\href}[2]{#2}\begingroup\raggedright\endgroup

\end{document}